\documentclass[epjCONF,columns]{svjour}
\usepackage{graphics}
\usepackage[varg]{txfonts} 
\usepackage[latin1]{inputenc}
\newcommand{\unitm}[1]{\,{\rm{#1}}}
\newcommand{\br}{{\cal B}}
\newcommand{\btoppk}{B^{+} \to p \bar p K^{+}}

\newcommand{\mpp}{M_{p \bar p} < 2.85}
\newcommand{\btoetack}{B^{\pm} \to \eta_{c} K^{\pm}}
\newcommand{\btojpsik}{B^{\pm} \to J/\psi K^{\pm}}
\newcommand{\etacpp}{\eta_{c} \to p \bar p}
\newcommand{\jpsipp}{J/\psi \to p \bar p}
\newcommand{\jpsi}{J/\psi }
\session-title{2011 Hadron Collider Physics symposium (HCP-2011),
     Paris, France, November 14-18 2011}
\begin{document}
\title{Measurements of the relative branching fractions of the $\btoppk$
decay channel including charmonium contributions}
\author{R. Cardinale\inst{1} \fnmsep \thanks{\email{roberta.cardinale@ge.infn.it}}
  on behalf of the LHCb Collaboration}
\institute{University of Genova and INFN Genova,
  Italy and CERN, Switzerland}
\abstract{
The study of the $B^{+}\to p \bar p K^{+}$ decay channel at LHCb
offers great opportunities to study different
aspects of the Standard Model and possibly Beyond Standard Model
physics. In particular it can be interesting not only for the
possibility to measure CP asymmetry but also to study possible
intermediate resonances.
The ratios of the branching fractions of the $B^{+}\to p \bar p
K^{+}$ decay channel, of the charmless component with $M_{p\bar p} <
2.85\unitm{GeV/}c^{2}$ and of the charmonium contribution $\eta_{c}$
relative to the $\jpsi$ are presented.} 
\maketitle

\section{Introduction}
\label{intro}
\begin{figure}[ht]
\centering
\resizebox{0.85\columnwidth}{!}{
\includegraphics{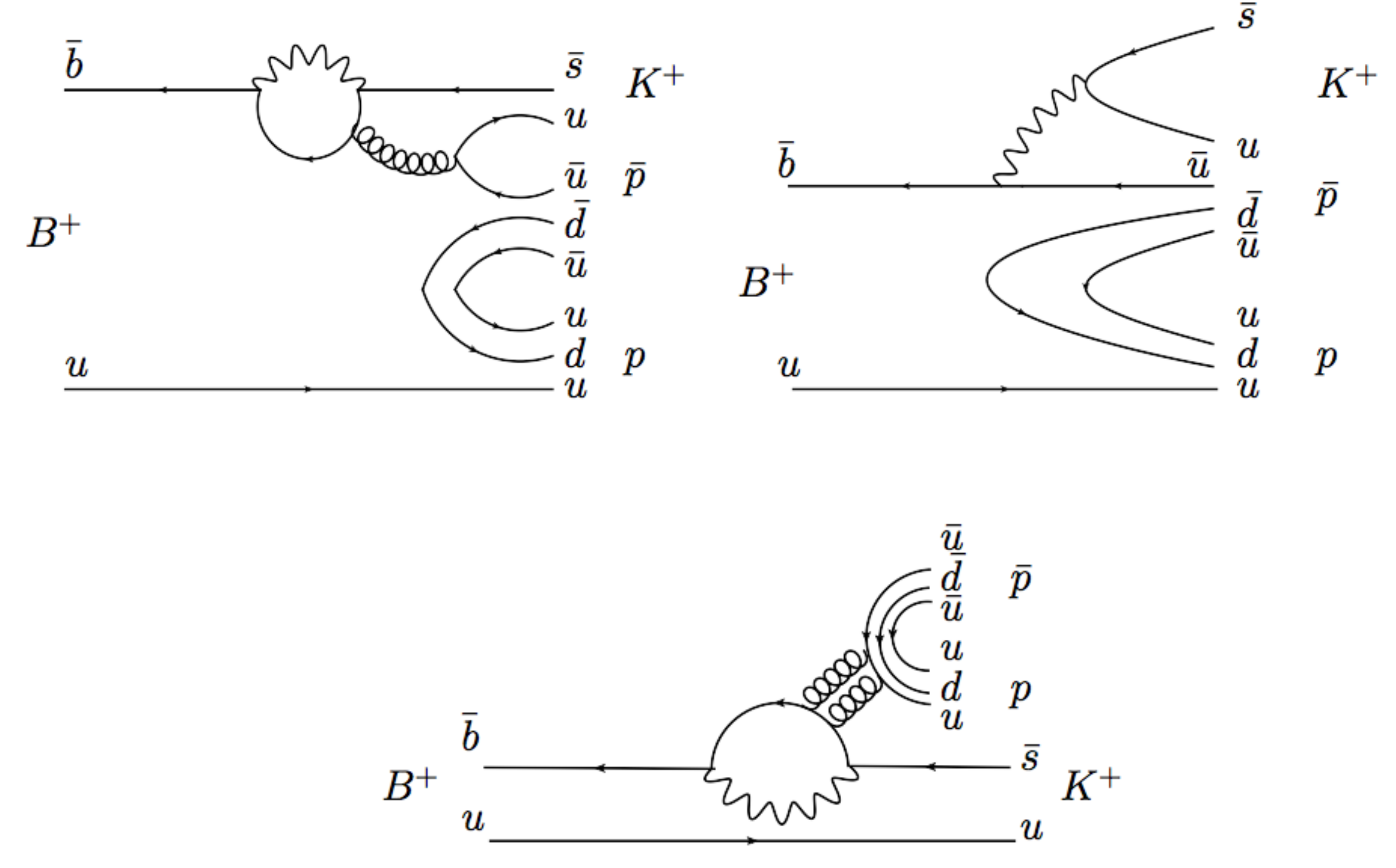}}
\caption{Feynman Diagrams: penguin (top left), doubly Cabibbo
  suppressed tree (top right), OZI suppressed penguin (bottom).} 
\label{fig:feyn}
\end{figure}

\begin{sloppypar}
The current knowledge on the $\btoppk$ decay is based on the
measurements performed at the
\mbox{$B$-factories}\cite{bi:BaBar,bi:Belle}.
Thanks to the high statistics
available
in the next years at LHCb, due to the large $b\bar b$ production
cross section, $\sigma_{b \bar b}\nolinebreak = (284 \pm 20 \pm
49)\mu\unitm{b}$~\cite{bi:lhcbcross} at a centre of mass energy of $\sqrt{s} = 7 \unitm{TeV}$, more precise
measurements can be performed and possibly new physics can be
revealed.\\LHCb is a single arm forward spectrometer optimised to perform precise CP
violation measurements and rare decays study in the heavy flavour sector.\\
Three-body baryonic $B$ decays are expected to proceed predominantly via $b\to s$ penguin diagrams or doubly Cabibbo suppressed tree diagrams~\cite{Cheng:2001tr} and as such new heavy virtual particles can give measurable departures from Standard Model expectations.
The baryon-antibaryon pair can also be produced by a pair of gluons in
OZI suppressed penguin diagrams, where gluonic resonances could be formed.
The three Feynman diagrams are shown in Fig.~\ref{fig:feyn}.
The interference between the penguin diagram and the doubly CKM
suppressed tree diagram can lead to a direct CP asymmetry.\\
Moreover, three-body baryonic decays offer a clean environment to study
intermediate states, as charmonium states, excited $\Lambda$ baryons and exotic states (glueballs
\cite{bi:glueball}, baryonium, pentaquarks).
In fact, the $p\bar p$ final state can be produced through a
resonant intermediate state, a $c\bar c$ state or through an unknown exotic state. 
At LHCb, beyond the study of well known charmonium states, 
it can be interesting to observe charmonium-like states such as $X(3872)$ or
new intermediate states in order to study their nature and properties.\\
Another interesting feature observed both at the BaBar and Belle
detectors in three body baryonic $B$
decays is the low mass $M_{p\bar p}$ enhancement. Several explanations
have been proposed. A study of the
distribution of events in the Dalitz plot can explain the origin of
this mass enhancement.~\cite{Rosner:2003bm}.\\
A preliminary study of this mode based on the
sample of events collected by the LHCb detector in the 2010 data
taking period is presented. The branching ratios of $B^{+}\to p \bar p
  K^{+}$, of the charmless component in the range 
$M_{p\bar p} < 2.85\unitm{GeV/c^{2}}$ and of the \mbox{$\btoetack$} decay
channel with
$\etacpp$ relative to the product of  branching fractions 
$\br_{J/\psi} = \br (\btojpsik) \times \br (\jpsipp)$, which is well
known~\cite{bi:PDG10}, have been measured.\\
The measurement relies only on the ratios of events, $N(\rm mode)$,
and efficiencies, $\epsilon_{\rm mode}$,
with respect to the reference mode:
\begin{equation}
\label{eq:br}
\frac{\cal{B}(\rm mode)}{\br (J/\psi)} = \frac{N(\rm mode)}{N(\jpsi)}\times
\frac{\epsilon_{\jpsi}}{\epsilon_{\rm mode}}\label{eq:main}
\end{equation}
where $N(J/\psi)=N_{B^{+} \to J/\psi
  (\to p \bar p) K^{+}}$~and~${\epsilon_{J/\psi}=\epsilon_{B^{+} \to
    J/\psi (\to p \bar p) K^{+}}}$
\end{sloppypar}

\section{Study of the $B^{+} \to p \bar p K^{+}$ selection}
Since the $b \bar b$ cross section is approximately $1\%$ of the total inelastic cross section and the interesting
channel branching ratios are of the order of $\mathcal{B} \sim
10^{-5}$-$10^{-6}$, it is extremely important to design a selection
with a high rejection of the background.
The selection strategy is based on a preselection with very loose
selection criteria. 
A multivariate analysis using the ``Boosted Decision
Tree'' algorithm of the TMVA package~\cite{bi:tmva} is used to optimise the
selection efficiency.
The variables used in the selection procedure exploit kinematic
and topological characteristics of $b$-hadron decays. Three
charged particles in the final state with large transverse momenta are
required. The three final state particles have to come from the same displaced secondary
vertex requiring a good vertex fit and a significant distance between the
two vertices. Moreover it is required a small maximum distance of
closest approach between the tracks and non zero impact parameters.
And the momenta direction of the
three daughters has to be compatible with the flight direction of the
$B$.
Furthermore the particle identification
information, provided by two RICH detectors in LHCb, is a fundamental
ingredient of the selection in order to
identify correctly protons and kaons candidates. 
A signal efficiency
of more than $70\%$ with respect to preselected events has been
obtained using the BDT algorithm with a signal to background ratio
of $S/B \sim 1$.

\section{Signal Extraction}
\begin{figure}
\centering
\resizebox{0.75\columnwidth}{!}{\includegraphics{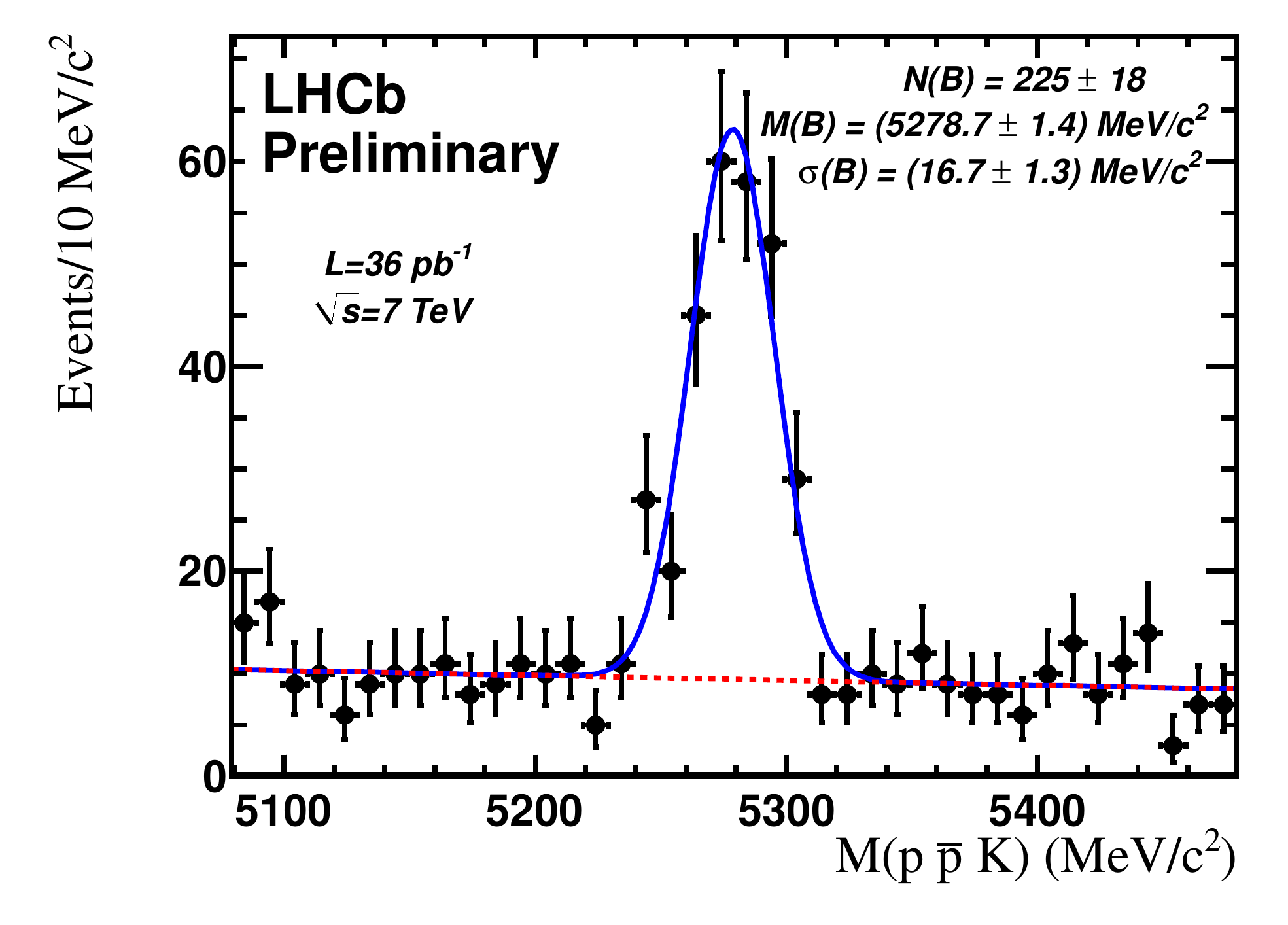}}
\caption{Mass distribution of all selected $\btoppk$ events. The continuous line shows the fit result, the dashed line shows the fitted background component.}
\label{fig:all}
\end{figure}

\begin{figure}
\centering
\resizebox{0.75\columnwidth}{!}{\includegraphics{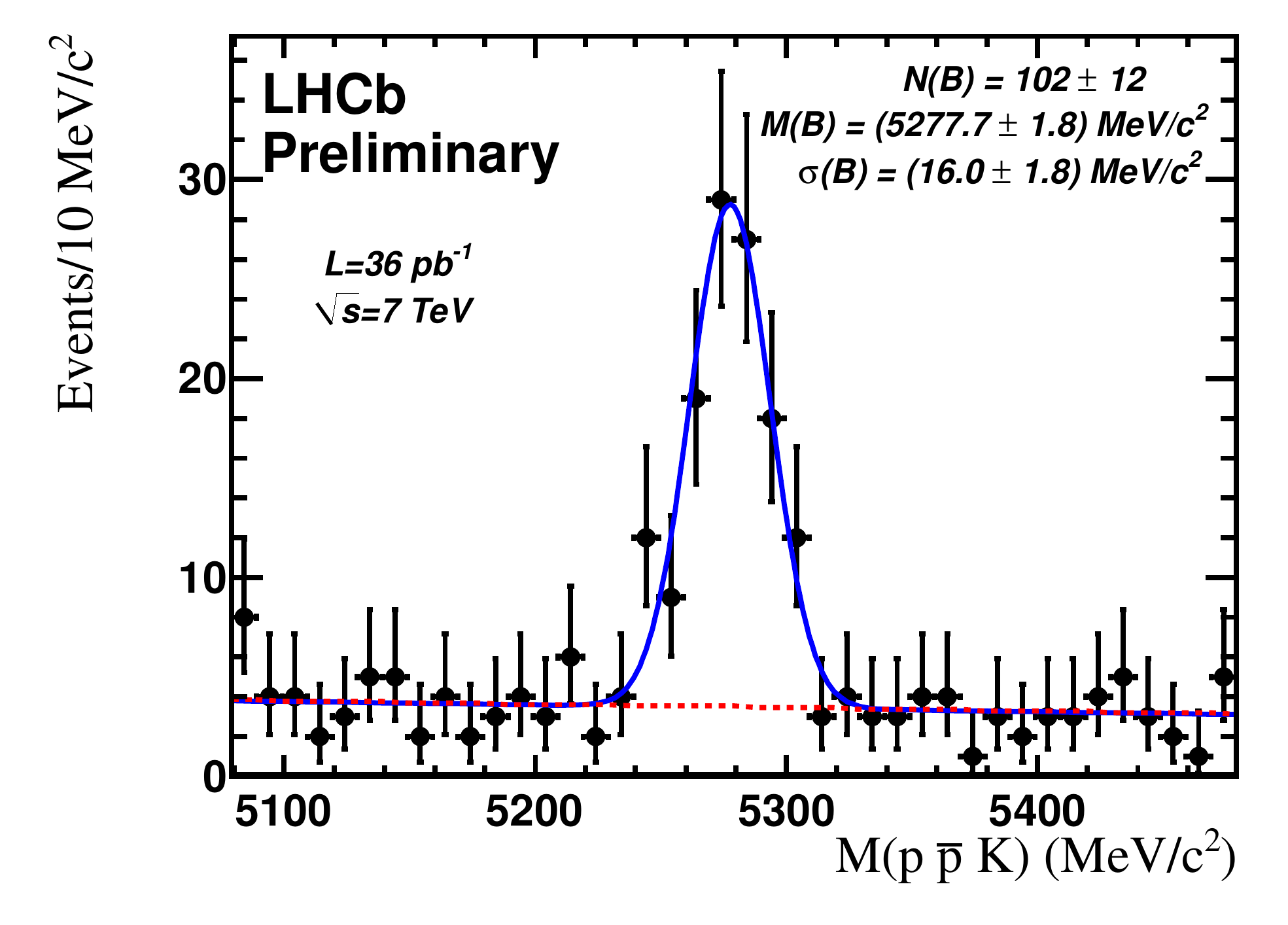}}
\caption{Mass distribution of the selected $\btoppk$ events for \mbox{$\mpp \unitm{GeV/}c^{2}$}. The continuous line shows the fit result, the dashed line shows the fitted background component.}
\label{fig:285}
\end{figure}

\begin{figure}
\centering
\resizebox{0.75\columnwidth}{!}{\includegraphics{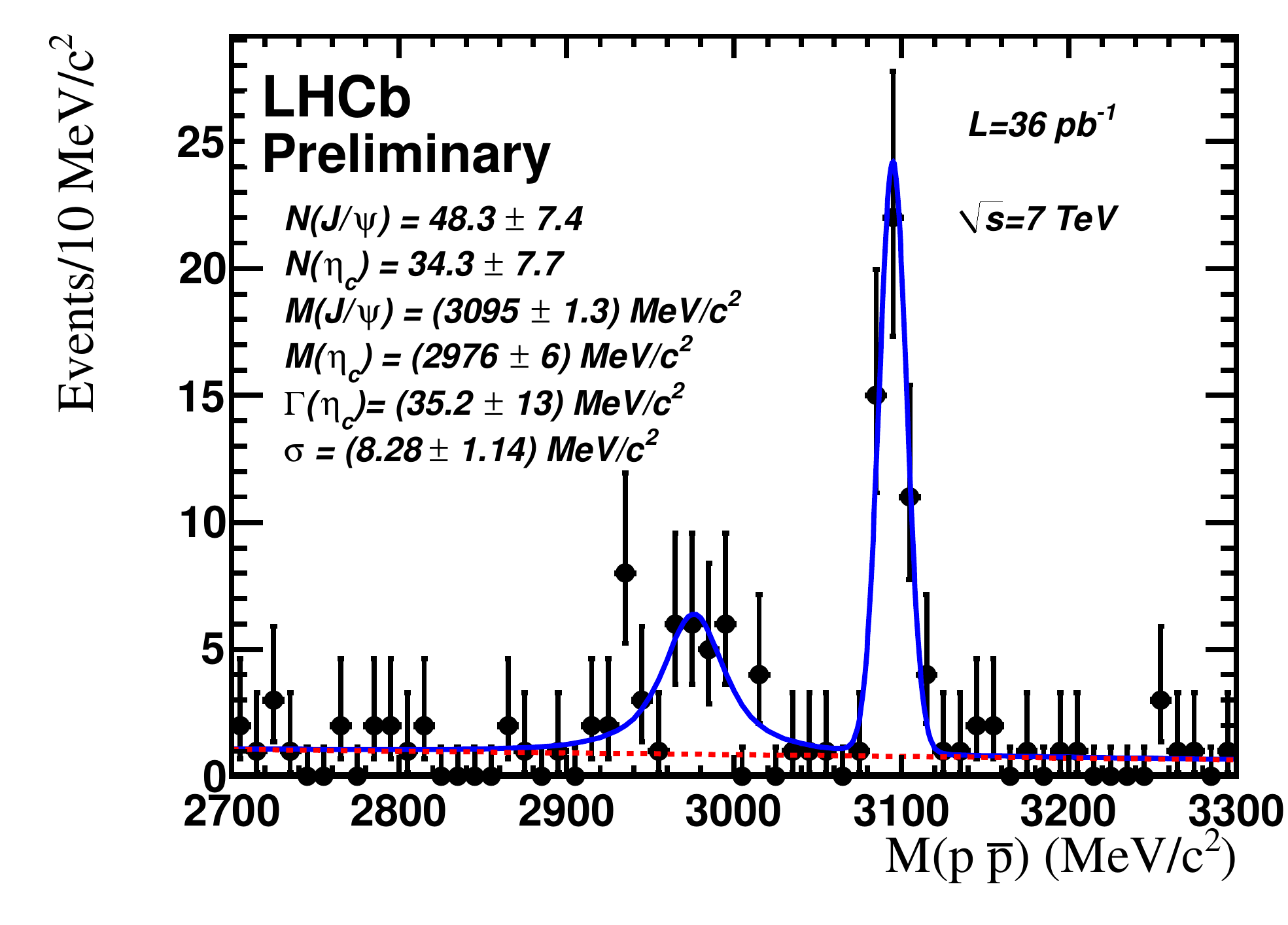}}
\caption{Mass distribution of $p\bar p$ in the \mbox{$\jpsi$-$\eta_{c}$}
  range for events having \mbox{$\vert M(p\bar p K)-M(B)\vert<50\unitm{MeV/c^2}$}. The solid line shows the result of the fit, the dashed line corresponds to the fitted background contribution.}
\label{fig:jpsietac}
\end{figure}

The number of $B^{+}\to p \bar p K^{+}$ signal events is
extracted from an unbinned maximum likelihood fit to the $M_{p\bar pK}$
invariant mass distribution of the selected events
(Fig.~\ref{fig:all}).
The fit procedure
has been applied also on the
subsample of events having $M_{p\bar p} < 2.85 \unitm{GeV/}c^{2}$ to
extract the number of events in the charmless region (Fig.~\ref{fig:285}). 
The signal probability density function has been parametrized with a
Gaussian function while for the background a linear function has been used.\\
For the $B^{+} \to
J/\psi (\to p \bar p) K^{+}$ and $B^{+} \to \eta_{c} (\to p \bar
p) K^{+}$, the yield has been extracted with a fit to
the $M_{p \bar p}$ distribution for the $B$ events having $M_{p\bar p K}$
invariant mass compatible within $2.5\sigma$ with the known $B^{+}$
mass (Fig.~\ref{fig:jpsietac}).
Since the $J/\psi$ meson has a natural width much smaller than the
detector resolution while the natural width of the $\eta_{c}$ is not
negligible, the $J/\psi$ has been parametrized with a Gaussian
function while for the
$\eta_{c}$ the convolution of a Breit-Wigner with a Gaussian function with the
same resolution of the $J/\psi$ has been used.

\section{Estimation of efficiencies}
\label{sec:eff}
The calculation of the branching fractions based on Eq.~\ref{eq:br}
requires the determination of the ratio of the efficiency for each
considered channel (all the events, events with $M_{p\bar p} < 2.85
\unitm{GeV/}c^{2}$ and $\eta_{c}$ events) with respect to the \mbox{$J/\psi$:
$\epsilon_{\rm mode}/\epsilon_{J/\psi}$}.\\
The overall efficiency ratios,
product of reconstruction, trigger and event selection efficiency
as a function of $M_{p \bar p}$ invariant mass have been calculated
using simulated events in order to correct for varying efficiency over
the $M_{p\bar p}$ range.\\The ratios with their uncertainties are reported in Tab.~\ref{tab:end}.

\section{Evaluation of systematics}
\label{sec:syst}
Since the final state for the considered modes and the reference one
is the same,
most of the systematic uncertainties cancel.\\
Possible systematic uncertainties have been considered on the event
yield determination and on the efficiency ratios.\\
The systematic uncertainty on the event yield extraction is estimated
using different fit ranges, different parametrisation for signal and
background. Half of the largest difference in the central value
obtained with different fit configurations is taken as systematic
uncertainty on the number of signal events.
The event yields for each mode and the corresponding systematic
uncertainties are listed in Tab.~\ref{tab:end}.\\ 
The uncertainty on the efficiency ratios described in
Section~\ref{sec:eff} is
considered as a systematic uncertainty.\\
Most data-simulation discrepancies are independent from the Dalitz
plot variables and cancel in the ratio. In some cases, in particular
for particle identification, the discrepancies are larger for low
momentum tracks. A conservative $10\%$ systematic uncertainty due to momentum
dependent data-simulation discrepancy on the  efficiency ratios has been assumed
for the total $\br(\btoppk)$ and for the charmless
$\br(\btoppk)_{\mpp\unitm{GeV/}c^{2}}$ branching ratios. 

\section{Branching ratio measurements}
Using the number of events extracted for each considered mode and the
corresponding ratios of efficiencies with respect to the $B^{+} \to
J/\psi (\to p \bar p) K^{+}$ the preliminary measurements of ratios
of branching fractions have been extracted and are reported in
Tab.~\ref{tab:bratio}.
\begin{table}[htbp]
\caption{Yields of $\btoppk$ events in each mode, ratio
  of efficiencies of each mode with respect to the $B^{+}\to\jpsi (\to
  p \bar p) K^{+}$.} 
\vskip 0.3cm
\centering
\begin{tabular}{|c|c|c|}
\hline
\hline
Mode &  Yield & $\epsilon/\epsilon_{\jpsi}$ \\
          &  $\pm$ stat $\pm$ syst
          & \\
\hline
\hline
$\jpsi$ & $48 \pm 7 \pm  2$& 1 \\
\hline
ALL  & $225 \pm 18 \pm 5$& $1.01 \pm 0.10$ \\
\hline
$\mpp\unitm{GeV/c^2}$ & $102 \pm 12 \pm 2$& $0.96\pm0.10$ \\ 
\hline
$\eta_{c}$ & $34 \pm 8 \pm 3$& $0.994 \pm0.003$ \\ 
\hline
\hline
\end{tabular}
\label{tab:end}
\end{table}

\begin{table}[htbp]
\caption{Ratios of branching fractions with statistical and systematic
  uncertainties for the different modes considered.} 
\vskip 0.3cm
\centering
\begin{tabular}{|c|c|}
\hline
\hline
Mode &  $\br$ ratio \\
          & LHCb \\
\hline
\hline
$\jpsi$ & 1 \\
\hline
ALL  & $4.6 \pm 0.6 \pm 0.5 $\\
\hline
$\mpp\unitm{GeV/c^2}$ & $2.21 \pm 0.41\pm0.24 $ \\ 
\hline
$\eta_{c}$ & $0.71 \pm 0.20 \pm 0.07$ \\ 
\hline
\hline
\end{tabular}
\label{tab:bratio}
\end{table}

\section{Conclusions and perspectives with the full 2011 data set}

The preliminary measurements of the ratios of branching fractions for
the $B^{+} \to p \bar p K^{+}$ decays have been reported. From the
measured ratios, using the known values of the $\br (B^{+}\to J/\psi
K^{+})$ and $\br(J/\psi \to p \bar p)$ the corresponding branching fractions have
been derived and reported in Tab.~\ref{tab:endbis}.

\begin{table}[htbp]
\caption{Derived LHCb preliminary branching ratios with statistical error,
  systematic uncertainty and the uncertainty on the
  knowledge of the $\br (B^{+}\to J/\psi (\to p \bar p) K^{+})$. All values are in
  units $10^{-6}$.}
\vskip 0.3cm
\centering
\begin{tabular}{|c|c|}
\hline
\hline
Mode &  LHCb  $\br$                    \\
          & (\small preliminary)  \\
\hline
\hline
ALL  &
$10.2 \pm 1.4 \pm 1.1 \pm 0.5$\\
\hline
$\mpp\unitm{GeV/c^2}$ &
$4.87 \pm 0.91\pm0.54 \pm 0.22 $\\
\hline
$\eta_{c}$ &
$1.57 \pm 0.43 \pm 0.15 \pm 0.07$\\ 
\hline
\hline
\end{tabular}
\label{tab:endbis}
\end{table}

\begin{table}[htbp]
\caption{Branching ratios measured at the $B$-factories. All values are in units $10^{-6}$.}
\vskip 0.3cm
\centering
\begin{tabular}{|c|c|c|}
\hline
\hline
Mode & BaBar & Belle\\
          & \cite{bi:BaBar} &\cite{bi:Belle}\\
\hline
\hline
ALL  & - & $10.76^{+0.36}_{-0.33} \pm 0.70$\\
\hline
$\mpp\unitm{GeV/c^2}$ & $5.3\pm0.4\pm0.3$ & $5.0^{+0.24}_{-0.22}\pm0.32$\\
\hline
$\eta_{c}$ & $1.8^{+0.3}_{-0.2}\pm0.2$ & $1.42\pm0.11^{+0.16}_{-0.20}$\\ 
\hline
\hline
\end{tabular}
\label{tab:bfactories}
\end{table}

The results are compatible with the average values reported in the PDG
and measured at the $B$-factories as reported in Tab.~\ref{tab:bfactories}, even with the limited statistics
used in this analysis. 

The current systematic uncertainty is largely of statistical origin, so
both statistic and systematic uncertainties will decrease as soon as the
available statistics increase. 

In 2011 LHCb has collected more than $1\unitm{fb^{-1}}$. The large
available statistics collected in 2011 is a factor almost $10$ more
with respect to the statistics collected at the $B$-factories. The
most accurate available
measurements have been performed by the BaBar and Belle Collaborations using
about $1000$ fully reconstructed exclusive decays. An improvement
on the precision of the branching fractions is expected in particular
for $\br(\btoetack)\times \br(\eta_{c} \to p \bar p)$. Moreover the
study of more rare charmonium contributions will be possible, such as
$\eta_{c}(2S)$ and $X(3872)$, observing or putting an upper limit on $\br (B^{+} \to c \bar c
K^{+}) \times \br(c\bar c \to p \bar p) \sim 10^{-7}$.

\end{document}